# Stress propagation in granular media:

## Breaking of any constitutive state equation relating local stresses together by a change of boundary conditions

**P. Evesque**

Lab MSSM, UMR 8579 CNRS, Ecole Centrale Paris
92295 Châtenay-Malabry, France, evesque@mssmat.ecp.fr

**Abstract:**

*The stress response of a granular assembly subject to different changes of boundary conditions is studied experimentally in order to define the stress propagation characteristics. These results demonstrate that no simple and single relationship between local stresses exists, which would be imposed by the local structure of the granular assembly only in general. On the contrary, it is demonstrated that these results are controlled by the boundary conditions themselves and that a tiny change of them may lead to strong variations in the incremememental-stress relation; furthermore, in other cases, these changes may generate large variations of the stress field and can allow to understand partly the fluctuations already observed in these media.*

---

## 1• Introduction :

Granular media have been the subject of many studies in recent years [1,2]. In particular, large interest has been carried on the prediction of the stress distribution in these materials and on their local fluctuations. From the mechanics point of view [3,4], it is in general well admitted that finding the stress distribution requires to know the boundary conditions and the rheological law of the materials, otherwise the system is indeterminate; more precisely, the number of stress continuity equations is not large enough compared to the number of stress components to raise the indeterminacy of the stress distribution so that it requires to associate to these equations a mass conservation equation together with the stress-strain behaviour.

It has been proposed recently an other way to approach this problem [1a]. It consists in forgetting the deformation parameter and in introducing a relationship between the local stresses [1a] of the kind $\sigma_{xx}=k\sigma_{zz}$. Doing so, the authors of [1a] have concluded to the possibility that stress propagates along lines.

This approach has been used in mechanics for which it corresponds to a state of perfect plasticity, which is known to lead to hyperbolic equations and hence to stress propagation along lines. However the domain of validity of such an approach in mechanics is limited to the case when perfect plasticity holds true, i.e. when the deformation path is not controlled by the stress increment. This hypothesis cannot be used for a static stable pile but only for marginally stable ones.

In this paper, we try to verify this restriction. So, the properties of stress propagation in a granular assembly is investigated experimentally, both at the scale of the whole medium and at a more "local" scale, (i.e. at a scale where the stress field is "homogeneous"). The "local" scale investigation uses a 3-D granular medium, on which one tries i) to verify experimentally that a relation between the local stresses does exist, ii) to demonstrate that this relation is imposed by the packing structure only, as it was assumed in [1a], iii) to demonstrate that this relation is never perturbed by a small increment of stress [1]; we will see however that the experiments deny the whole set of assertions in general.

When concerned with the global scheme of propagation, the experimental investigation is made on 2-D assemblies of rods with different shapes. One of the simplest method used in this case consists in perturbing the stress distribution of a rectangular medium by applying a small force at a given location and measuring the stress response at an other location. The dependence of this stress response upon the amplitude of the force and upon the force location has been determined; it turns out that the major parameter controlling the response is the gauge characteristics; the harder it is the larger the response, and when the gauge is soft, the response is zero. So, it results from this that the stress propagation direction is controlled by the gauge: if the gauge is soft, the stress tries and avoids the location of the gauge; on the contrary, if it is hard, the stress is attracted by it in a given ratio which depends on the relative position of the gauge and of the location where the force is applied. This demonstrates that the propagation direction is neither an intrinsic characteristic of the internal structure of the pile only, nor of the initial stress state. At this stage, the stress propagation appears as if it was a non local problem.

Other experimental results are also reported for 2-D assemblies. They all lead to conclude that no direction of stress propagation is fixed by the pile structure alone, nor by the pile structure and by the initial stress state. On the contrary, they prove that stress propagation depends greatly on boundary conditions and on their perturbation. We analyse and interpret these experimental results within this framework and find a good agreement with data.

This work is divided into two parts. The first one investigates the response of a 2-D granular assembly when applying some given small force at some location. The second one is concerned with 3-D experiments.

## 2• 2-D granular medium with hexagonal packing:

This section focuses on the mechanics response of a 2-D rectangular granular medium made of rods of equal diameter and having an internal triangular structure to different changes of boundary conditions. Consequences about the way the stress propagates into the pile are drawn.

### *2•1 Experimental set-up:*

The experimental set-up is sketched on Fig. 1. It is made of a rigid structure on which a 2-D granular medium is laid. The rigid structure is made of two parallel vertical U-shape structures fixed together; their inner height and length were 16cm and 16.3cm respectively and the distance separating the two parallel U is 3.5cm. The granular medium is made of parallel duralumin cylinders of 5mm diameter, 6cm long and $m_o$=3.1g mass, with their axes perpendicular to the U-shape structure; its internal structure is dense and is approximately triangular (hexagonal); the basis is horizontal.

The length of the rectangular pile was 16.3cm and its height H was varied from $H_m$=3 cm to a maximum height $H_M$ = 11.7cm, which corresponds to 7 and 27 layers of rods respectively.

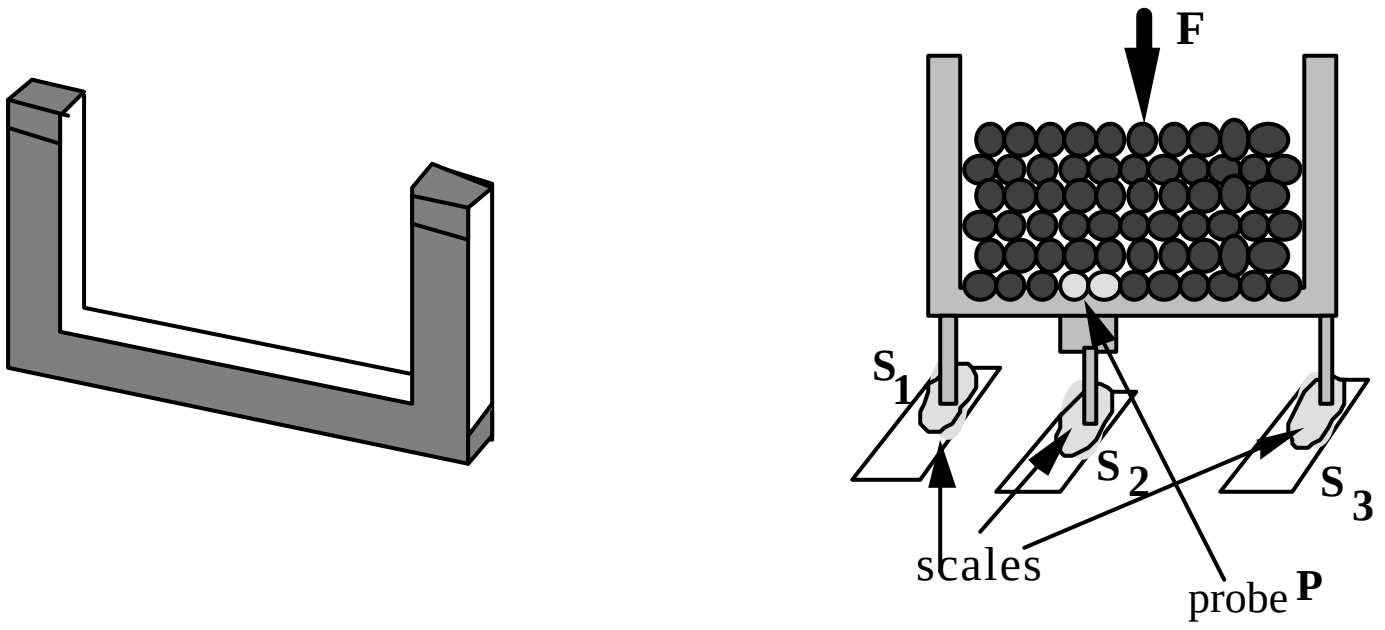

**Figure 1: Experimental set-up:**
**Fig. 1a**: the structure on which the granular medium lays is made of two parallel U-shaped structure with a void in between them

**Fig. 1b**: rods are laid on the structure; they are 5mm diameter and 6cm long; probe rods P are 3cm long and 5mm diameter. This probe is carried by a structure which passes in the space between the two U and which lays on the scales $S_2$. The set-up lays on two other scales $S_1$ and $S_3$. Some additionnal weight F may be applied at different points of the top layer; this induces a change of the weight measured by each scales. Two of these scales are hard, the third one is much softer.

As the structure consists of two parallel U separated by l=3.5cm, it allows to insert at the bottom of the pile some probe P which does not touch the structure, but which is directly supported by a support which passes through the double U without touching it; the support reposes on a spring balance labelled $S_2$ (see Fig. 1) so that the vertical load it supported can be measured. The probe P is made by a single cylinder whose length (3cm) is smaller than l=3.5cm so that it does not touch the U; however other probes have been used; which were made by gluing together few (1 to 5) cylinders of 5mm diameter and 3cm long, so that different areas could be checked. This set-up allows then to measure the vertical stress at this location. It is also worth mentioning that these probes have just the same structure as a small part of the bottom row of the pile so that they come just replacing the initial rods (which have been put away before insertion) and without disturbing its internal structure; so, these probes feel the same stress than would have supported the initial rods. The balance $S_2$ can be moved up and down by a special carriage to adjust the position of the probe P..

Each end of the U structure reposes also on scales, which are labelled $S_1$ and $S_3$. On the top of the pile, different loads can be added in different position and the effect on the probe stress can be measured. During these experiments, we used three different scales, two of them were hard the other one much softer. We could also adjust the height of each support.

The experiment consists in measuring the loads $W_1$, $W_2$ and $W_3$ carried by each scales $S_1$, $S_2$ and $S_3$ when different loads $W_0$ are applied on top of the pile, at different locations.

### *2•2 Experimental result: response of the rectangular pile to an applied localized overload*

The experiment consists in considering a given configuration stress distribution, which will be characterised by the three weights $W_{1o}$, $W_{2o}$ and $W_{3o}$ carried by each scales $S_1$, $S_2$ and $S_3$ , to apply some overload $W_o(x)$ at some location x on the top layer of the pile and to measure the variation $\delta W_1(W_o(x))$, $\delta W_2(W_o(x))$ and $\delta W_3(W_o(x))$ this overload generates. The to vary the point of application of the extra force. The set-up is this one of Fig. 1b. It uses three scales $S_1$, $S_2$ & $S_3$, one of which is much softer than the two others, with a stiffness K=7.33µm/g =0.733mm/N. It has been used either as $S_2$ or $S_3$.

Label $S_i$ the signals from the scales i (i=1 to 3). It can be shown that the initial $S_2$ value can take any value in a given range $\{0,W_{2max(0)}\}$, when no overload is imposed to the pile; $W_{2max(0)}$ depends on the pile height and increases when an overload F is applied on top of the pile, in a given range of location; i.e. $W_{2max(F)}$ follows the law : $W_{2max(F)}=W_{2max(o)}+F$.

In Fig. 2, experimental values of $S_{2(F)}=S_{2(F=0)}$ are reported as a function of the position of the overload F, for different positions of the soft scales (Figs. 2a and 2b) and for different values of the overloads. Fig. 2a (2b) has been obtained with $S_2$ (S3) as the soft scale . Experimental data of Fig. 2 show that $S_{2(F)} = S_{2(F=0)} +F$ is not true, since the signal $S_2$ from the scale $S_2$ does not depend on F in Fig. 2a or does depend on F in Fig. 2b but with a proportionality coefficient which i) is not 1 and ii) which depends on the location where the force F is applied.

Furthermore, experiments have shown that i) the variations $\Delta S_i = S_{i(F)}-S_{i(F=0)}= S_i-S_{io}$ of $S_i$ does not depend i) on $S_{io}$, ii) on the number of layers constituting the rectangular pile and iii) on the probe size. Experiments have shown also that $\Delta S_i =S_{i(F)}-S_{i(F=o)} =S_i-S_{io}$ varies linearly with F when F is applied at the same location and varies linearly with the position of application if F is kept constant. It can be also shown that $S_1+S_2+S_3-F=S_{1o}+S_{2o}+S_{3o}$ and depends linearly on the number of layers of rods the pile is made of.

• *interpretation:*

A simple way to understand these results is to consider the pile as rigid. It is submitted to a force F located in x and to three forces $S_1$, $S_2$ and $S_3$ applied by the scales in $x_1$=0, $x_2$ and $x_3$=L. Furthermore, as one of the balance is soft and as the two others are rigid, the force applied by the soft scales shall remain constant. Label M the mass of the pile and of the structure which holds it. Equilibrium of forces and of angular momentum imply:

$$F+Mg=S_1+S_2+S_3 \tag{1a}$$

$$Fx+Mgx_G=S_2x_2+S_3L \tag{1b}$$

**Figure 2:** A rectangular pile is carried in three points by three scales (see Fig. 1b). The variation of response of each scales $S_1$-$S_{1o}$,(squares), $S_2$-$S_{2o}$,(triangles), $S_3$-$S_{3o}$ (losanges) when an overload (M=200g or 50g), is plotted as a function of the overload position on the top of the rectangular pile. The probe is linked to $S_2$ and is made either of 1 or 5 rods, (see caption title). The variation of the weight does not depend on the initial values of $S_{1o}$, $S_{2o}$, $S_{3o}$, nor on the probe size, nor on the pile height (21 or 27 rods layers), but on the position of the less rigid scales.

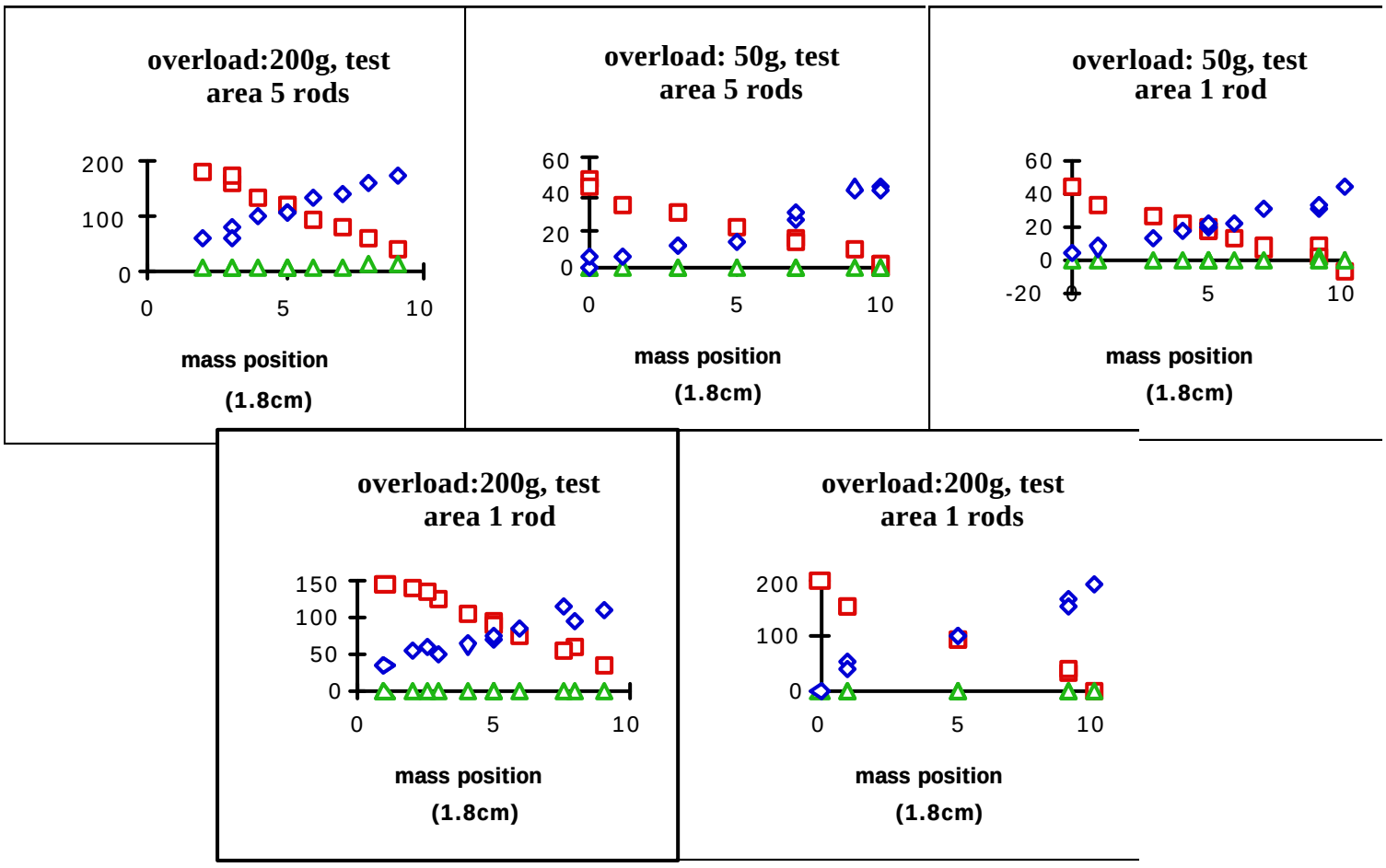

**Fig. 2a:** $S_2$ is the softest-spring balance; this fixes the response of $S_2$ to be constant.

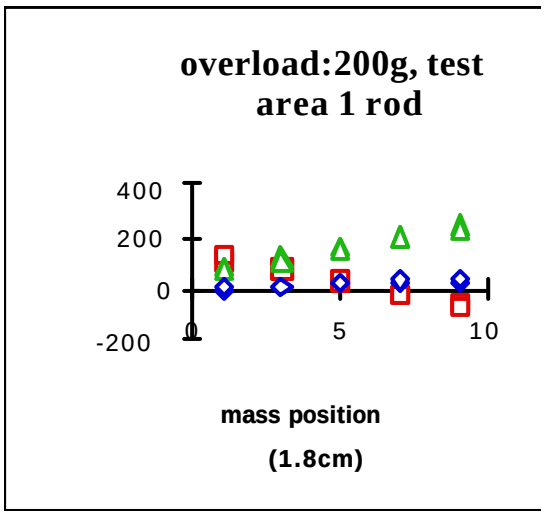

**Fig. 2b:** $S_3$ is the softest-spring balance; this fixes the response of $S_3$ to be constant.

where $x_G$ is the absicsse of the center of gravity. As one knows x, F, M, $x_G$, $x_2$, and the force applied by the soft scales (either $S_2$ or $S_3$), one gets if $S_2$ is the soft scales:

$$S_1=[F(L-x)+Mg(L-x_G)-S_2(L-x_2)]/L \quad (2a)$$
$$S_2=cste \quad (2b)$$
$$S_3=(Fx+Mgx_G-S_2x_2)/L \quad (2c)$$

whose solution is:

$$\Delta S_1=F(L-x)/L \quad (3a)$$
$$\Delta S_2=0 \quad (3b)$$
$$\Delta S_3=Fx/L \quad (3c)$$

or, one gets if $S_3$ is the soft scales:

$$S_1=[F(x_2-x)+Mg(x_2-x_G)+S_3(L-x_2)]/x_2 \quad (4a)$$
$$S_2=(Fx+Mgx_G-S_3L)/x_2 \quad (4a)$$
$$S_3=c^{ste} \quad (4a)$$

whose solution is:

$$\Delta S_1=F(x_2-x)/x_2 \quad (5a)$$
$$\Delta S_2=Fx/x_2 \quad (5b)$$
$$\Delta S_3=0 \quad (5c)$$

These behaviours are observed respectively in Figs. 2a and 2b.

### *2•3 Consequences about stress propagation in the pile as assumed in ref. [1]:*

It has been argued recently [1a] that stress shall propagate along stress lines in a sandpile. This result has been obtained in 2d by remarking that there are three unknown stresses in two dimensions $\sigma_{xx}$, $\sigma_{xy}$ and $\sigma_{yy}$ and that there are only two stress continuity equations so that one needs to introduce a third equation to close the mathematical problem. One has then tried and introduced a linear relation between two of these stress fields. Doing so, one finds that the stress shall propagate along two straight lines whose orientations are symmetric compared to vertical and inclined at $\theta$ from this orientation (cf. Fig. 3). So, according to this approach, and applying a small overload F at a given point x on top of a rectangular pile, shall lead to the generation of two equal loads at the bottom of the pile at locations $x \pm h\, tg\theta$. This is not what it is observed in the experiment. So:

• *no simple stress propagation along lines in general*

The experimental results reported here denies the validity of this point of view and shows that the propagation of stress is controlled by the boundary conditions.

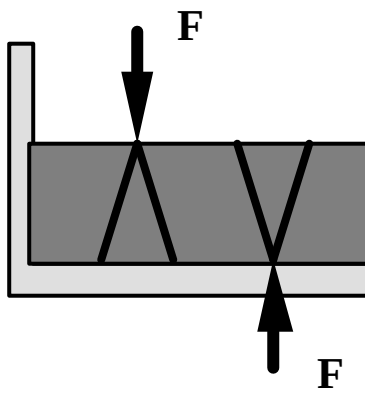

**Figure 3: Fig. 3a:** Stress propagation "according to ref [1a]".

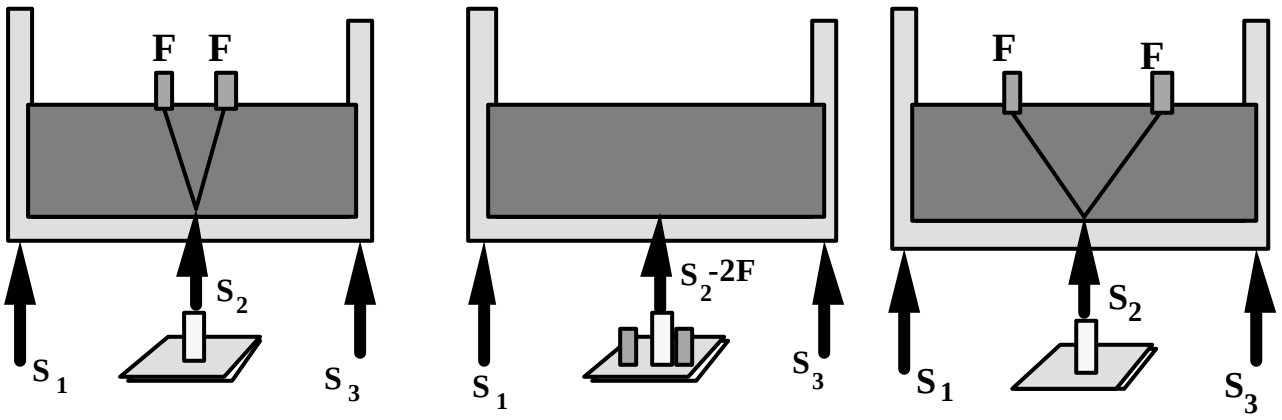

**Fig. 3b:** how to get a pseudo stress propagation "along two lines".

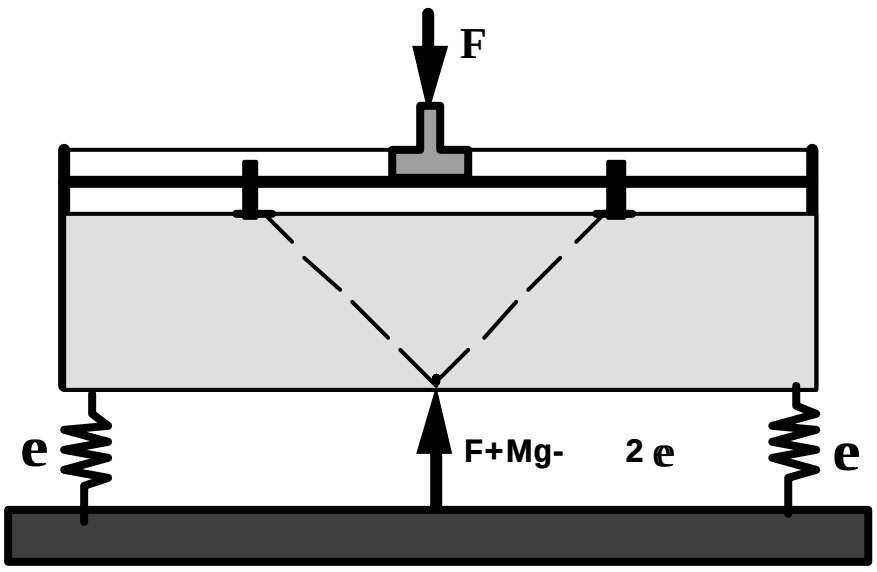

**Fig. 3c:** a set-up which might lead to a stress which propagates "along two lines": the pile is contained in a rigid container; the 2 soft springs ensure the equilibrium of the system, but most of the weight is carried by a rigid balance located in the middle which applies the force $F+Mg-2\varepsilon$. Increasing the force F on top applies on the two pistons a force F/2 which is transmitted directly to the scales below, through the granular medium.

For instance everything occurs as if point $S_2$ was repealing the stress, when $S_2$ is the soft scales. On the other hand, if $S_2$ is one of the two rigid scales, point $S_2$ attracts always part of the added stress whatever the position x! It is also worth noting that not only the boundary conditions influence the stress distribution, but also those which control the mechanics of the structure on which the pile is fixed, since the scales $S_1$ and $S_3$ do not act directly on the boundary of the pile, but act on them through the 2-U structure.

An other experimental result which is worth commenting concerns the force one can apply upward at any location on the bottom of the pile. When applying the approach developed in [1a], one finds that the small overload shall propagate along two directions and should remain localised along two lines (Fig. 3a). This leads to concludes that whatever small and positive the overload, it should generates the upward motion of this part of the pile which is concerned with. This is obviously not observed experimentally and demonstrates that this approach is not valid in general and shall be handled with more care.

The real problem which is raised with approach of ref. [1a] comes from the difficulty to combine an approach of rigid body for which local internal stress distribution does not govern the mechanics since the system is rigid which implies in turn that no mechanical energy can be defined.

On the other hand, the classical approach is based on a thermodynamics approach which requires to introduce the two conjugated variables i.e. the stress and the strain; it requires also to know the strain stress energetic coupling. In this case, one can show that stress- and strain- continuity equations added to the local strain-stress behaviour is enough to solve the problem as a function of the boundary conditions.

In other words, the approach of ref. [1] is only possible when the boundary conditions are changed in such a way to verify some rules which are in agreement with the state equation of stress used; otherwise, it will need to introduce an uncontrolled change on the state equation so that the new stress distribution respects the change of boundary conditions. But the prediction of the change requires the knowledge of the initial stress distribution and to solve the whole problem using also the deformation.

• *how to get a stress which "seems" to propagate along lines as in ref [1a]*

Does this mean that one can never observe a process of stress transmission similar to that one found in [1a], (i.e. as in Fig. 3a). As a matter of fact this is not difficult to get: take a rectangular pile, with the soft scale in the middle ($S_2$). Load the top layer with two equal masses m symmetrically located compared to $S_2$. Then at the same time remove these two masses m (with F=mg) and put them directly on the scale $S_2$. One observes no variation of the signal from $S_1$ , $S_2$ and $S_3$; this implies that the force applied to the probe has varied by 2F during the removal of the two masses, so that the stress 2F "propagates in the desired directions" towards the two masses and converge there. The trick here is that the change of boundary conditions has been chosen carefully to be compatible with the desired propagation: Move the two masses back on two other locations symmetric compared to $S_2$ and you get two new "directions of propagation". The sketch of this experiment is given in Fig. 3b; it has been tested experimentally and the result confirm this analysis.

Nevertheless, if one knows where the forces converge when using such variations of boundary conditions; no one knows what are the real change of stress distribution deep in the pile.

In the same way, Fig. 3c proposes a set-up which shall generate a stress propagation which looks like straight-line: applying an increment of force $\delta F$ on the top central part will apply an increment of force $\delta F/2$ on each piston; the stability of the system of forces imposes that this generates a force $-\delta F$ at the middle of the bottom of the pile in order to counterbalance the added load. One can even chose the pseudo direction of propagation of the stresses by choosing the length separating the two pistons.

*However, it is worth mentioning that in all these experiments, one can speak only of a pseudo "direction of propagation" since no real measure of the internal-stress field variation is made.*

So, a sum-up of the results obtained in this section 2 can be made as follows. The 2-D experimental pile, which has been under investigation here, has a triangular internal structure. This means in peculiar that its density is quite high so that its dilatancy is important prior than a deformation occurs. In this case, one could think that the pile is nearly rigid and that its deformation will be always negligible compared to the deformation of the soft-spring balance. This imposes that the soft-spring force remains constant during an experiment, which changes the position of the extra-force.

This demonstrates that the approach proposed in [1] is never valid when the pile can be considered as rigid.

• *Fragile matter versus fragile constitutive law:*

This demonstrates that the approach proposed in [1] is never valid when the pile can be considered as rigid, i.e. when it is harder than the gauges. So, there are two possibilities. i) In the first case, the constitutive law is valid; this indicates that the system obeys perfect plasticity so that the medium is at the limit of yielding and will yield (or break) indefinitely when a small stress perturbation is imposed. ii) In the second case, the medium is either rigid or can accommodate small deformations; in this case this is not the medium nor the matter but the constitutive law itself which is fragile [5,6].

• *Remark: can one conclude really that stress does not propagate along lines with this experiment?*

One of the referees remarked that since theory reported in [1] respects the basic requirements of equilibrium (zero force and zero torque), it certainly cannot lead to results in contradiction with Eq. (1) and hence with the experimental data.

In other words, theory in [1] imposes a set of differential equations which includes the constitutive equation, the solution of which depends on the boundary conditions. *So, approach in [1] does not impose the boundary conditions; so, the stress field it predicts shall depend on these boundaries and shall vary with these boundaries*.

The answer to this objection is the following: there is no guarantee that the set of differential equations has always a solution for any set of boundary conditions. In particular it can be shown that there is no pair of solutions obeying i) the constitutive law ($\sigma_{xx}=k\sigma_{zz}$) , ii) the set of differential equation and iii) the pair of boundary conditions of the present experiment (M. Cates, referee report).

## 3• Incremental results on 3-D granular medium:

In ref. [1,5,6], it is assumed the existence of some relationship between the stresses; an example of such a relation is $\sigma_{xx}=k\sigma_{zz}$. It is also assumed that this relation depends on the material history and is engraved in the material as the main information.

In this section, we argue in a different way: of course one may find a granular medium for which such a relation stands, so that $\sigma_{xx}=k\sigma_{zz}$ for instance; but what we want to stress here is that one may perform on this granular medium different stress increments $\delta\sigma_{xx}$ and $\delta\sigma_{zz}$ which do not follow this relation in its incremental form: $\delta\sigma_{xx}\neq k\delta\sigma_{zz}$.

Different experimental set-ups have been used to investigate the stress-strain behaviours of granular materials and their sensitivity to history; among them one finds the Casagrande box, the shear annulus box, the axi-symmetric triaxial set-up, the true triaxial one, the $1\gamma$–$2\varepsilon$ apparatus.... We report the reader to any textbook on soil mechanics for this subject. We focus here with what is found using a triaxial cell for sake of simplicity.

Typical results obtained with a triaxial cell driven at constant pressure p are reported on Fig. 4. They have been obtained from the same sand with different initial density [7]; but they concern only initially isotropic packing. More information about the understanding of these results can be found in [7] or in textbook of soil mechanics. It is obvious from this experiment that $\delta q\neq\delta p/p$, so that the relationship assumed in ref [1] is not satisfied in this test procedure. On the contrary this test preserves the direction of principal axes. So, it is run in the context of some FPA principle similar to that one assumed in some cases in ref. [1b].

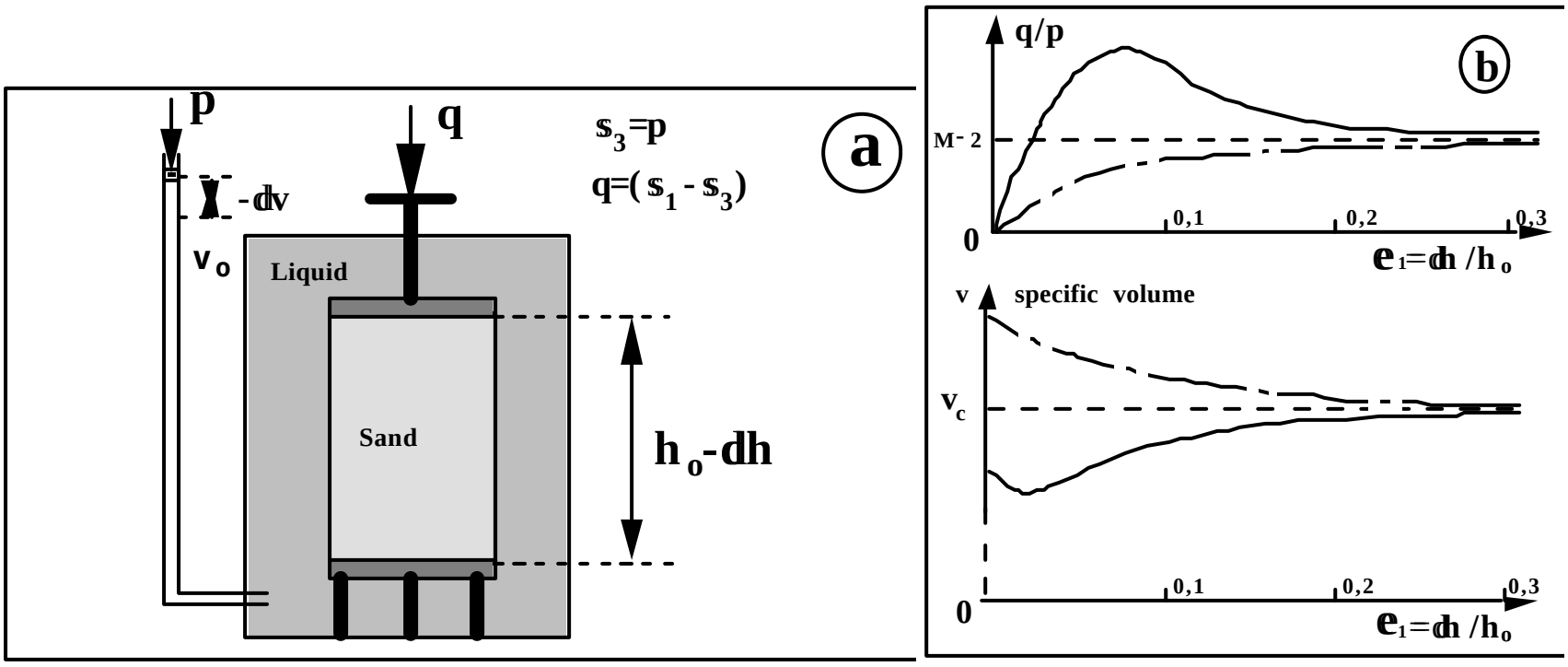

**Figure 4:** A 3-D granular medium made of rigid grains can deform under stress.

**Fig. 4 a:** a typical axisymmetic triaxial test set-up consists of a plastic cylindrical bag which contains the granular medium; it is immersed in a container filled with water at pressure $p=\sigma_3$ and maintained in between two vertical pistons which applies a variable vertical overload q.

**Fig. 4 b:** The mechanical behaviour is summed up by the knowledge of the three following parameters $\sigma_3$, the deviatoric stress $q=(\sigma_1-\sigma_3)$ which characterises the shearing force, the vertical strain $\varepsilon_1$ and the volumetric strain $\varepsilon_v=\varepsilon_1+2\varepsilon_3$ or the specific volume v. Here, typical results obtained with a triaxial cell, when $\sigma_3$ remains constant. One obtains that the asymptotic value $v_c$ depends on p, but not on the initial specific volume $v_o$ and that the asymptotic value M does not depend on p and $v_o$. The transient behaviour depends on both $v_o$ and p

But it is also known that one may use the same granular medium in the same initial state of density,... and use an other set-up such as the $1\gamma$–$2\varepsilon$ set up, or the shear box to impose to this granular medium to follow a different stress path which does not preserve the direction of the principal axes. It is known that one would get results similar to those of Fig. 4. So one cannot conclude that the conservation of a given law is ensured; everything depends on the apparatus, and/or on the variations of boundary conditions one imposes.

At last, it is worth mentioning that one can use the triaxial cell set-up to impose to work at constant radius and constant direction of principal stress; doing so, one gets an "oedometric" test. In this case, and if one considers tests for which q is kept increasing continuously (and hence p), one reaches a stationary evolution after a while. This stationary regime is such that the q/p ratio is kept constant while q and p increase and this test follows the law: $q=k'_{Jaky}p$ ; ($p=\sigma_3$ in this notation). The constant $k_{Jaky}$ is called the Jaky constant and is related approximately to the friction coefficient $tg\varphi$ by the law obtained from a best fit from experimental results: $k'_{Jaky}=\sin\varphi/(1-\sin\varphi)$. We have proposed a theoretical demonstration of this law recently; it is based on the analysis of the deformation path, which contradicts the hypotheses of [1,5,6]. So, one can conclude that such a test works under both the FPA assumption and the q=kp relation, *but it works due to the existence of deformation.* Furthermore the *preservation of this law fails when q is decreased*: after a large transient behaviour, the discharge rate reaches the law: $q/p= 1-tg^2(\pi/4+\varphi/2)$. As a last remark on this oedometer test, one may find an experimental justification of the assumption made by Janssen for silo calculations through the Jaky's law; nevertheless, one shall keep in mind that this approximation is only true during the charge of the silo and is not true during the silo discharge. It is also worth noting that this relation assumes the wall friction to be zero, otherwise the stress distribution is no longer uniform in a horizontal slice [10]. In the same way, and for the same reason, it is at last worth mentioning that applying a local stress at a given point will also destroy in general the validity of this law, at least in its incremental form: $\delta q\neq k'_{Jaky}\delta p$ .

## 4• Conclusion

This paper investigates the response of granular materials to different loads at both a global point of view (section 2) and at a local one (section 3); it focuses on their dependencies upon variations of boundary conditions.

This study does not agree with the approach proposed in ref [1,5,6] who assumes i) that strain is not a right parameter and ii) that it exists 1 (or 2) intrinsic closure equation(s) in 2-D (or 3-D) relating stresses between them. This is due to the fact that variations of boundary conditions can be arbitrary in the present study, while it cannot be arbitrary in [1,5,6].

It is worth noting that most of the experimental results of section 2 was not concerned with any deformation problem, but only with a global problem of equilibrium. In this case one is faced to solve the global system of equations of equilibrium, which consists of the equations of i) conservation of momentum and ii) of conservation of angular momentum; and this is enough to know "how stress propagates to outside the

material». Experimental results of sections 2.2 are in agreement with this viewpoint and they demonstrate that varying the boundary conditions can be done in such a way that it violates any closure equation as assumed in [1] without violating the real stability of the system, so that approach [1] fails and so that experimental results of section 2 deny completely the point of view developed in [1] for which there shall exist one or two additional simple rheological laws relating the stresses together to close the mathematical problem and find the stress state. It denies also “fragile” nature of the material as supposed in [5,6] because it is not concerned with the pile deformation.

It is also worth mentioning that our approach of solving the global mechanical equilibrium of the system to interpret the experimental data of section 2 does not imply that the stress distribution inside the material is known; it does not implies also that the medium is rigid indeed, since same results would have been obtained with an elastic piece of metal for which it is known that the stress does not propagate along lines.

In section 3, emphasis has been carried upon the local stress-strain behaviour of a 3-D granular medium in order to demonstrate that deformation is a right parameter of the mechanics of theses systems. So, from section 3, it is clear i) that experimental stress-strain laws do exist and ii) that any stress increment can be imposed to a granular material, whatever its stress state is, as far as the medium is not at limit of yielding. Theses results are then in direct contradictions with the hypothesis of paper [1] which assumes the proportionality between the stress increment ratios and the stress ratios: $\delta\sigma_{ij}/\delta\sigma_{kl} = \sigma_{ij}/\sigma_{kl}$ at any local position.

Turning now to the incremental point of view, it is straightforward to conclude that the knowledge of the stress-strain behaviour of the granular material in each point allows attributing energy to any distribution of stress. So, one may apply a principle of minimum energy to know which stress distribution will be achieved from an initial condition and according to local stress equilibrium and local mass conservation. This leads to the adequate number of equations to solve the stress distribution. This method is used classically in the case of elastic materials and leads to the elasticity theory of stress. This is also the classical approach used in soil mechanics to predict the stress distribution in granular medium; but in this case the local stress-strain behaviour does not obey to perfect elasticity and one uses more complex rheological laws, such as elasto-plasticity and/or its incremental formulation.

As a final remark, and in order to convince the still dubious reader, let ask the following question: We were able to understand the experimental results of section 2 by introducing the stiffness of the three balances. Would it be possible to understand these results when considering the three scales as rigid? Since in this case there are three unknown loads $S_i$ (i=1,2 & 3), and only two equilibrium equations, the problem is indeterminate; each scales can take any value in a given range; and the load on each scale can vary drastically as a function of time, depending on slight fluctuations,...! This would be exactly what would occur in a granular medium if this one were rigid, in a more complex scheme of course.

The first written version of this paper, the experimental set-up and the experiment were presented for the first time in the poster session of the Powders & Grains 97 meeting (Durham, NC, USA, May 1997). They have been presented also at the Institut of theoretical Physics, University of California at Santa Barbara, during the NATO programm "Jamming & Rheology" (UCSB Sept. 97). The paper and its modified versions were rejected by *J. de Physique France* and by *Physica* D. this version is very near the first version. It does not report a set of experimental results however, since it has nothing to do with the main topics of the paper; also, it takes into account few referee remarks.

*Sans la liberté de critiquer et de contester, il n'y a ni science, ni scientifiques, ni éloge flatteur mais que des "grandeurs d'établissement"[11].*

## Appendix: Other experiments on 2-d ordered triangular piles

In the next, the apparatus of Fig.1 has been used to characterise the limit of stability of the pile; the experiment consists in determining the range of stress one can applied to the probe as a function of the shape and the height of the pile and/or for different loading.

### *A-1 maximum localised force that can be applied to the bottom of a rectangular pile:*

In this experiment the size of the probe was one rod, and we applied different loading using the soft-spring balance as $S_2$ . So, varying the height z of this balance using the special carriage allows to varies the vertical stress applied by the scales to the probe and hence to the pile, according to the relation $W_2=K(z-z_o)$, where K is the stiffness of the scales, z the vertical axis and x the horizontal one; K=7.33µm/g =0.733mm/N.

It has been observed that the load $W_2$ that can be applied to the probe can take any value in a given continuous range. This one starts from 0; its maximum $W_{2max}(N,W_o)$ depends on the number of horizontal layers N constituting the rectangular pile and on the overload $W_O$ applied to the pile, when $W_o$ is located in a given range of horizontal position x not too far from the $x_p$ location of the probe (i.e. $|x-x_p|=ND\cos\alpha$, with $\alpha=60°$, D=5mm is the grain diameter). It has been observed also sometimes that the probe can be "stuck" into the pile, so that it stays in place when the scales is removed (i.e. when $W_2=0$); in this case, this probe seems to "levitate". This is the proof of the existence of some pure arching effect allowed by some friction between the probe and the pile.

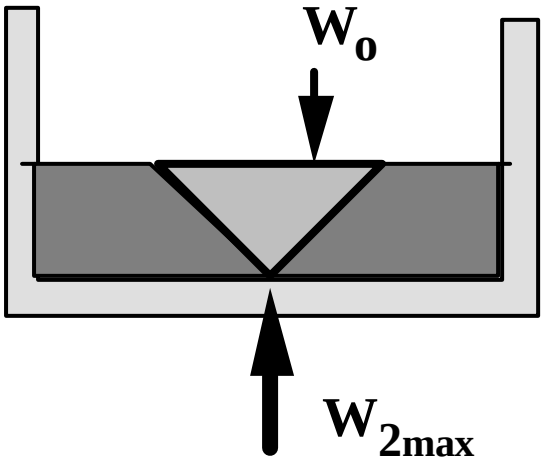

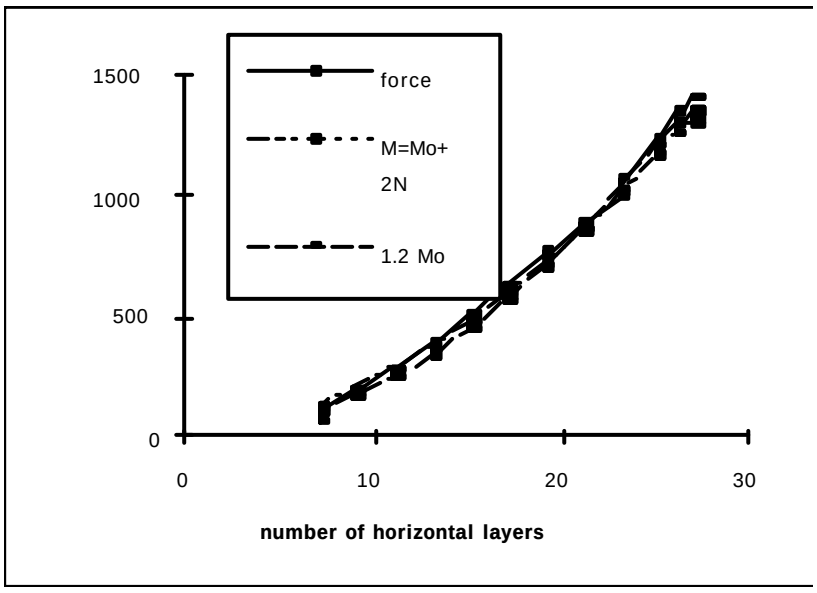

**Figure 5: Fig. 5a:** sketch of the experiment with a rectangular pile subject to different loads. The pile is stable as far as the bottom force does not exceed a given value $W_{2max}$, which depends on the pile height N and on the overload $W_o$. When $W_2$ reaches $W_{2max}$ , a finite deformation occurs in two zones inclined at 30° from vertical and the triangle moves up. The motion is a fraction of the grain diameter and leads to a diminution of $W_2$ .

**Fig. 5b:** Maximum vertical upward force which can be applied on the basis of the one-cylinder probe by the soft scales, as a function of the number N of layers containing the pile. Black squares are experimental; the two other curves are best fits obtained either by taking account of the mass of the triangle as $F= 1.2M_Og$ with $M_o=N(N+1)m_o/2$ is the mass of the rods pertaining to the equilateral triangle or as F=Mg with $M=(N(N+1)/2 +2N)m_o$ is the mass of the triangle which contains also the localised-deformation bands (of 1-grain width). $m_o$ is the mass of a single cylinder (3.1g).

If one starts increasing $W_2$ from $W_2$=0, one observes first that the internal structure remains invariable. In a second step, a deformation becomes visible, even if it remains quite small and corresponds to an upward motion of the probe of a tiny fraction of grain diameter; this deformation is caused by a redistribution of the contacts since the pile structure is not perfectly triangular with six contacts per grain, but is rather unperfected and has 4 contacts per grain. This is true till $W_2$ reaches $W_{2max}$. But when $W_2$ overpasses slightly $W_{2max}$ , it decreases abruptly and a macroscopic upward motion occurs, whose amplitude is a fraction of D (see next paragraph). It has been observed experimentally that $W_{2max}$ fluctuates from experiment to experiment within few percents.

• *Deformation*

The deformation (Fig. 5a) occurs along two sliding zones oriented at 30° from the vertical; they start from the probe grain. The width of these localisation zones fluctuates from experiment to experiment but is of one- or two-grain diameter about in general. These two bands define five zones: the two bands where deformation occurs, and three zones where no (or quite tiny) deformation exists; these last three zones are the equilateral triangle and the left and right sides of the pile. As already mentioned, the internal structure of the packing located in the three zones which are not deformed are not disturbed and remains triangular, so that the deformation seems to occur as the upward motion of a rigid equilateral triangle laying on its tip at a macroscopic scale. This tip is also the probe.

Focusing now on what is occurring inside the two bands where the deformation is localised, one sees that the deformation starts occurring by a uniform series of vertical sliding at each point of contact between rods such that their tangent is parallel to vertical. So, this deformation is the combination of a series of vertical dislocations localised along two inclined lines oriented at 30° from vertical; this means in turn that these localisation bands cannot be viewed as a single dislocation but has a more complex behaviour. During the upward motion the triangle expands laterally in order that its grains remain in contact with the grains remained immovable. It is worth noticing that this process is reversible: diminishing $W_2$ allows restoring the initial packing structure.

• *Maximum force:*

$W_{2max}$ has been measured as a function of N and $W_o$ . As already mentioned, $W_{2max}$ (N, $W_o$ ) fluctuates from experiment to experiment when keeping constant N and $W_o$ ; the amplitude of fluctuations is about 100g when $W_{2max}$ is 1400g (i.e. N=27) and $W_o$=0; but it is less for smaller pile height. Furthermore, using a statistical treatment of the data concerning different measures with the same height of pile but with different overloads $W_o$ allows to demonstrate that $W_{2max}$ is simply the sum of the contribution linked to the equilateral triangle and of the overload $W_o$ sitting on this triangle. So one gets:

$W_{2max}(N, W_o) = W_{2max}(N, W_o=0)+W_o$ if $W_o$ is applied above the triangle which is supported by the probe and

$W_{2max}$ (N, $W_o$ )= $W_{2max}$ (N, $W_o$=0) when $W_o$ is applied outside.

So the right quantity to measure is $W_{2max}$(N)=$W_{2max}$(N, $W_o$) -$W_o$ when $W_o$ is applied above the triangle. Averaging over $W_o$ of the {$W_{2max}$(N)=$W_{2max}$(N, $W_o$) -$W_o$} data has then been performed and plotted as a function of N (cf. Fig. 5b). One expects that this quantity corresponds approximately to the weight of the triangle, which is gN(N+1)$m_o$/2 , with the gravity acceleration g. However, this mass is slightly smaller than that one which corresponds to $W_{2max}$(N). So, better fits of the experimental data are obtained by introducing either i) a renormalisation coefficient k=1.2 which is aimed at taking account of the friction between the grains, ( i.e. $W_{2max}$=1.2 gN[N+1]$m_o$/2 ), or ii) by renormalising the mass of the triangle by including in it the mass of two localised zones of thickness one grain so that M becomes M=$m_o$[($N^2$+3N)/2). These best fits are plotted in Fig. 5b for comparison and agree rather well with the experimental data; so it is difficult to select one of them. The experimental fluctuations may be understood as generated by the fluctuations of the thickness of the sliding zones.

It is also worth mentioning the partial compatibility of these results with the approach proposed by Rowe [8] for the deformation of a pile: in the present case, the rectangular pile cannot dilate horizontally and it is in its densest structure; so horizontal motion of grain is impossible and upward motions of them are required. So, according to the Rowe's approach, one is then faced to find the orientation of two lines of grains on top of each other such as the orientation of the contacts between the grains pertaining to each lines are vertical or can be broken by a vertical motion. Introducing as another condition the fact that the lattice structure is triangular with a horizontal axis allows concluding that the only possible orientation of grain lines is the two other axes of the triangular lattice, which are at 30° compared to the vertical. The shape of the deformation agrees then with the Rowe's approach.

However, in Rowe's method [8], the force needed to perform the deformation is calculated by assuming some friction coefficient between the grains; this leads to a correction of the force-needed proportional to the mass of the triangle. So, this approach is compatible with fit i (i.e. $W_{2max}$=1.2$M_o$g=1.2N[N+1]$m_o$g/2 ); It is not compatible with fit ii (i.e. $W_{2max}$=($N^2$+3N)$m_o$g/2).

### *A-2• maximum localised force that can be applied to the bottom of a trapeze pile:*

The shape of the pile studied in this section is a trapeze with variable height whose bottom is horizontal and occupies the basement of the 2-U structure, so that the basement length is L=16.3cm, and corresponds to $N_o$=32 rods; it has three free surfaces, two of which are inclined at $\alpha$/2=30° from vertical and the last one is horizontal, see Fig. 6a. When the height H= 0.5 L tg$\alpha$, the pile is an equilateral triangle. The first and second rows of rods of this triangle contains both $N_o$ horizontal layers to stabilise the pile in a configuration compatible with the fact that L=16.3 cm is not an integer of the grain diameter D=0.5cm. It results from this that the number of horizontal layers contained in the triangle is No+1=33 and that its mass is [($N_o$+2)($N_o$+1)-2]$m_o$/2= 1736 g.

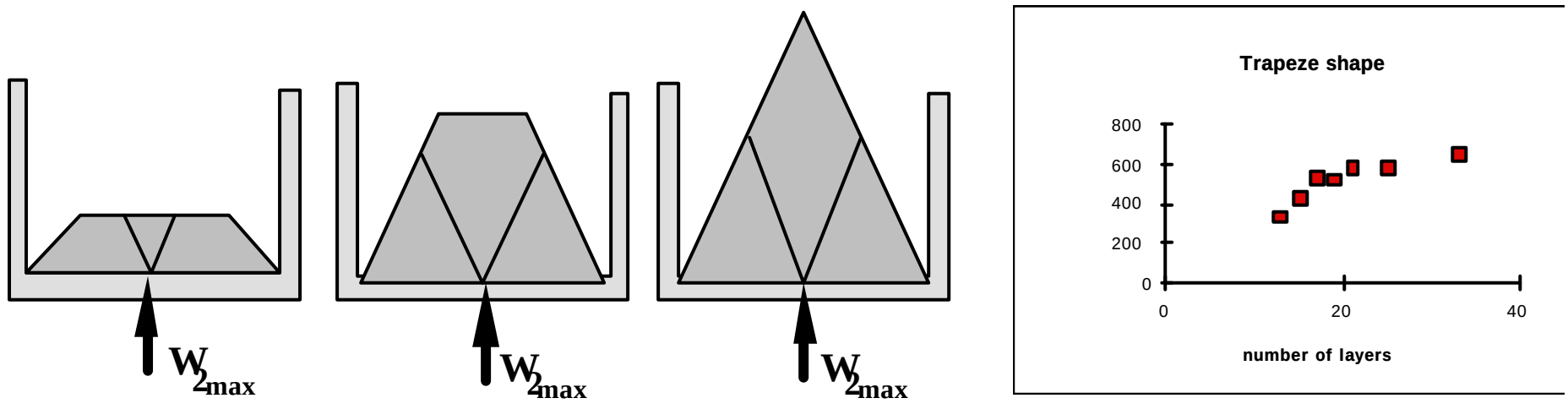

**Figure 6: Fig. 6a**: The different shapes of the studied trapeze pile.
**Fig. 6b:** Maximum vertical overload $W_{2max}$ which generates the trapeze breaking as a function of its height N measured in horizontal grain layers.
The piles have always the same basement length (32 rods). For N larger than 15, the pile breaks generating important side flows. For N=13, an overload F=200g on top of the pile can be applied, but F=500g generates the breakage. For N=15, an overload F=200g provokes the breakage.

When the pile is a trapeze, its structure is obtained by removing the up-most tip of the pile from the triangle. The probe is made of a single half rod and has been located just in the middle of the basement line of rods, so that the problem keeps the same symmetry whatever the load applied to the probe.

It is a remarkable feature that one can applied to this probe a vertical upward force $W_2$ which can take any value in the range {0, $W_{2max}$}. $W_{2max}$ is found to depend on the number of horizontal layers N; its variations are reported in Fig. 6b , in the range N=13-33. For N values smaller than 13, one expects results similar to those obtained in Fig. 6b, since side effects shall be negligible since they are limited to a region touching the corner of the U-shape structure and extending over a length equal to N grains. This is what is found experimentally: the difference between $W_{2max}$ measured in Fig. 6b and 2c for N=13 and 15 is either 50g or 100g. But as far as N approaches or overpasses the value $N=N_o/2=16$, side effects are no more negligible, which explains the difference between Figs. 2b and 3b. One remarks that $W_{2max}$ corresponding to the full triangle structure ($W_{2max}/g$=640g) is less than the total mass of the two equilateral head-to-tail triangles supported by the probe i.e. $M=[(N_o/2+1)(N_o/2+2)+N_o/2(N_o/2+1)]m_o/2=(N_o/2+1)(N_o+2)m_o/2=896g$ . This is the proof that the deformation process is different from that one of Figs. 2b or 3a. This is confirmed by the experiment, since one observes that the pile blows up laterally when $W_{2max}(N_o=33)$ is reached, which was not the case for the rectangular pile. Such lateral explosion is observed as soon as N>15 and it becomes quite important for N>21.

In the same way, one can applied an overloads F=200g on top of the pile (see Fig. 1) for pile lower than $N_o$=13. Doing so and applying an overload F=500g makes the pile N=13 unstable.

At last, it is worth mentioning that the value $W_{2max}$(N=32)=640g corresponds approximately to 40% of the mass of the whole triangle! This is a huge proportion of the pile which is carried by a single grain just in the middle of the pile! This result has to be kept in mind for comparison with any approach trying to compute a stress distribution at the bottom of a triangular pile: i) according to the location where the stress is applied, there is no reason that the symmetry of the stress distribution be broken; ii) this result shows that the stress distribution can be quite singular at a given location and can contain quasi Dirac distribution functions. In some sense, this result disagrees with ref. [1b].

At last, it is worth noting that the above behaviours are characteristic of a 2d ordered and dense packing since the structure of the deformation is mainly influenced by the packing structure and the direction of contact surfaces between the cylinder. So these results would be quite different if structural disorder of the packing was added, or in 3d. Experiments have been carried on the last case and much smaller load has been found indicating that the penetration mechanism looks much more as that one observed with a penetrometer.

The electronic arXiv.org version of this paper has been settled during a stay at the Kavli Institute of Theoretical Physics of the University of California at Santa Barbara (KITP-UCSB), in june 2005, supported in part by the National Science Fundation under Grant n° PHY99-07949.

*Poudres & Grains* can be found at :
http://www.mssmat.ecp.fr/rubrique.php3?id_rubrique=402